\documentclass[a4paper]{article}
\usepackage{RR}
\RRNo{6340}
\usepackage{t1enc}
\usepackage{hyperref}
\usepackage{graphicx}
\RRdate{Septembre 2007}
\RRauthor{% les auteurs
Cristian Ruz
  \thanks{The described research work and results have been obtained 
during Cristian Ruz's internship at INRIA Sophia-Antipolis, 
co-funded by a CONICYT/INRIA scholarship 
and the INRIA ARC Automan from December 2006 to April 2007}%
  \and
Fran\c{c}oise Baude\thanks[sfn]{OASIS team - INRIA, I3S, Universit\'e de Nice Sophia-Antipolis, CNRS}
\and
Virginie Legrand-Contes%
\thanksref{sfn}}
\authorhead{Ruz \& Baude \& Legrand-Contes}
\RRtitle{Towards Grid monitoring and deployment in Jade, using ProActive}
\RRetitle{Towards Grid monitoring and deployment in Jade, using ProActive}
\titlehead{Towards Grid monitoring and deployment in Jade, using ProActive}
\RRresume{Ce document montre comment utiliser le style RR.sty.
Pour en savoir plus, consulter le fichier RR.dvi ou RR.pdf
D\'efinissez toujours les commandes avant utilisation.
}
\RRabstract{This document describes our current effort to gridify
Jade, a java-based environment for the autonomic
management of clustered J2EE application servers, developed
in the INRIA SARDES research team.
Towards this objective, we use the java ProActive grid technology. We
first present some of the challenges to turn such an autonomic
management system initially dedicated to distributed applications
running on clusters of machines, into one that can provide
self-management capabilities to large-scale systems, i.e. deployed
on grid infrastructures. This leads us to a brief state of the
art on grid monitoring systems.
Then, we recall the architecture of Jade, and consequently
propose to reorganize it in a potentially more scalable way.
Practical experiments pertain to the use of the grid
deployment feature offered by ProActive to easily conduct 
the deployment of the Jade system or its revised version on any sort of grid.}
\RRresume{Ce document d\'ecrit nos efforts actuels pour gridifier Jade, un environnement Java destin\'e \`a l'administration des serveurs d'application J2EE clusteris\'es (r\'epliqu\'es), d\'evelopp\'e dans l'\'equipe de recherche INRIA SARDES.
Afin d'atteindre cet objectif, nous utilisons la technologie Java ProActive. Dans un premier temps, nous pr\'esentons quelques \'etapes \`a franchir  afin de transformer un tel syst\`eme d'administration initialement d\'edi\'e aux applications r\'eparties sur des clusters de machines, en un syst\`eme qui peut fournir des fonctions d'administration autonomique aux syst\`emes \`a large \'echelle, c.a.d. d\'eploy\'e sur des infrastructures de grille. Dans cet
objectif,  nous pr\'esentons un bref \'etat de l'art sur les syst\`emes de supervision des syst\`emes distribu\'es sur la grille.
Dans un deuxi\`eme temps, nous rappelons l'architecture de Jade, et nous proposons en cons\'equence de la r\'eorganiser de fa\c{c}on \`a ce qu'elle passe potentiellement  plus \`a l'\'echelle. Enfin, des exp\'erimentations pratiques se r\'ef\`erent \`a la fonctionnalit\'e de d\'eploiement offerte par ProActive dans le but de faciliter le d\'eploiement de Jade ou de ses versions r\'evis\'ees sur n'importe quel type de grille.}
\RRmotcle{supervision de grilles, deploiement sur grille, ProActive,
administration autonome de syst\`emes r\'epartis }
\RRkeyword{grid monitoring, grid deployment, ProActive, autonomic management
of distributed systems}
\RRprojet{Automan/OASIS}
\RRtheme{\THCom \THNum} % cas de 5 themes
\URSophia % pour ceux qui sont au Sud.
\begin{document}
\makeRR   % cas d'un rapport de recherche
%% \makeRT % cas d'un rapport technique.
%% a partir d'ici, chacun fait comme il le souhaite
\section{Introduction}

This report describes some research effort and results obtained in the
context of the 
%work done during the internship titled 
%``Using ProActive for provisioning grid nodes in the context of J2EE servers replication".
%This work is developed in the context of the 
AutoMan project
\url{http://sardes.inrialpes.fr/research/AutoMan/}, an INRIA funded
research collaboration between
OASIS, SARDES teams, and LSD team from UPM, Spain.

The objectives of AutoMan are to study the autonomic management of grid-based enterprise services.
In this context, Jade, a framework for autonomic management of distributed applications,
developed by the SARDES team  is considered. Specifically, one
of the Automan purposes is to evaluate the burden for such an autonomic
management system to extend its applicability and
usability from a clustered to a grid set of machines.

Jade is designed using the Fractal component model, and implemented as
a Java application using Julia and Fractal RMI for supporting remote interactions between the Jade
components that may be distributed on different machines of the cluster.
In order to be autonomically managed, a legacy application, for example a 
J2EE application server, must be wrapped inside a Fractal
component. Thus, a common interface  is provided to manage these legacy
applications from the Jade framework \cite{BouchICAC05,Cluster}.
Jade deploys the modules that constitute the 
architecture of the  wrapped legacy
application and  mirrors this architecture in order to further monitor it.  
%This way it %
% allowing its management 
As Jade needs to deploy and remove legacy modules, applications have to be deployed as
bundles on OSGi platforms. The Jade system itself is deployed as OSGi bundles. Thus, this
requires the previous deployment of OSGi gateways. 
The currently used support for Jade is Oscar,
\url{http://oscar.objectweb.org} an open source implementation of OSGi.
Indeed, the use of OSGi permits very easily the autonomic (re)deployment of (replicated)
modules in case of failure or poor performances, by simply
uploading OSGi bundles on available OSGi gateways already
running the Jade system.

%Wrapped applications can be deployed as OSGi bundles. For this purpose, the Oscar implementation is used.
%In fact, the Jade framework itself can be deployed as an OSGi bundle.

%As part of the research directions followed in the AutoMan project, are:
%   \begin{itemize}
%      \item Scalable replication of grid-based entreprise services.
%      \item Scalable monitoring of a set of grid nodes.
%      \item Scalable deployment of Jade over a grid infrastructure.
%   \end{itemize}

\bigskip
Enabling Jade to apply its autonomic management strategies 
at a grid level instead of just a cluster of machines has
the following underlying aims: 
\begin{itemize}
\item to give better performance scalability of
the deployed legacy application, as the number of replica can
be increased as will (not taking into consideration
other aspects such as database replication which may not well
apply at a grid scale -- this is an other matter, also studied in Automan), 
\item to increase flexibility, as all replica
are not constrained to be located on the same cluster and
can be migrated or restarted on other
machines to e.g. better balance the overall load of the clusters
or desktop  machines forming the grid.
\end{itemize}

These aims pose requirements on the management system, i.e. Jade:
\begin{itemize}
\item capability to deploy itself and the applicative modules
on any computing environment.
For this, the ProActive grid middleware developed within the OASIS team
\cite{BBCCHMQ05,CDD:CMST06} can be relevant as it supports a completely open, configurable deployment model:
through XML-based deployment descriptors, the user can abstract away
protocols, job submission systems,
and launch or get access to a process (a native process or a JVM).

\item capability to enable Jade-level and application-level communication
between any participating entity, whatever its location.
Here again, the use of ProActive could be relevant. Both inter ProActive runtimes
and ProActive active objects or software components communications
take grid constraints into consideration (latency hiding
through asynchronous with future remote single or multipoint
method invocation, on-demand securisation of communications -- authentication,
non repudiation, encryption). The constraints in order to get immediate
advantage of such properties (at least within Jade itself) would be to rely on
 ProActive instead of Fractal RMI for the interactions
between the Fractal Jade components.

\item capability to share and distribute the management related
operations and associated messaging according to the way
the managed system is deployed (amount and location of the replicas).

This means that the architecture of the Jade system itself must
be flexible enough, to be replicated at a grid
scale.
\cite{BouchICAC05} previously devised a replicated version of the Jade
architecture for the purpose of making Jade itself a self-repairing
system. Consequently, the effort was put on the needed protocols
for the various Jade replicated modules (Manager component) to
always maintain a coherent copy of the global state of the managed
system and globally synchronize their operations.
Here, our aim is different because  we aim at proposing a more
scalable and grid-enabled version of the
Jade architecture to better adhere to the effective
deployment of the managed application.
Indeed, Jade may be in charge of a big number
of widely distributed applicative entities or components at once.
So, the Jade monitoring operations (get sensed data, trigger
actions, ...) must also be scalable and grid aware.
%How to make this capability be grid-aware, scalable ? 
This is why we first studied the state of the art about grid monitoring: with
the aim to get insight on how  a monitoring system should be
designed and operate onto either the set of
grid resources or the set of applications deployed
on this set. 
\end{itemize}

%The first monthes of the internship included a revision of the state of the art related to grid monitoring,
%and a learning stage on ProActive and the architecture of Jade. 
This report first presents a short state of the art we collected about Grid Monitoring, with the intention to get insight onto how to make Jade more grid-aware, i.e., more scalable.
The following sections show some issues about the architecture of Jade,
describes how ProActive can be used to deploy it on a grid,
and finally suggests a hierarchical organisation of Jade
for it to better suit scalability requirements.
%Finally some on-going and future work is described.

\newpage
\section{State of the art on Grid monitoring}

In this section, we briefly present and analyse some of the relevant works
regarding architectures for providing scalable and efficient Grid Monitoring.

\subsection{Grid Information Services for Distributed Resources Sharing}
\subsubsection{Paper reference}

         K.Czajkowski, S.Fitzgerald, I.Foster, C.Kesselman, Grid Information Services for Distributed Resources Sharing, 10th IEEE Int. Symposium on High-Performance Distributed Computing (HPDC-10), 2001
\subsubsection{Brief summary of the project described in the paper}
This work describes an architecture to build an information service on top of different Virtual Organizations (VOs)\footnote{A VO can be understood as a
set of virtually aggregated resources forming a grid}, 
in order to share its resources with other VOs. 
Grid applications can benefit from these services to know which resources are available, since this availability could be dynamic.

Some of the services required by grid applications include: 
   \begin{itemize}
      \item Service discovery service, to know about new resources available.
      \item Application adaptation agent, to monitor and modify application behavior.
      \item Superscheduler, for routing requests to the 'best' resources (where 'best' must be defined).
      \item Replica selection service, to request 'best' copies of replicated resources.
      \item Troubleshooting service, to monitor and look for anormalous behavior.
      \item Performance diagnosis tools, when an anormalous behavior can be identified.
   \end{itemize}

The proposed architecture  is composed of {\em Information Providers}, and {\em Aggregate directory services}.
{\em Information providers} form a common, neutral infrastructure providing access to dynamic information about grid entities.
{\em Aggregate directories} obtain information through the providers, and answer queries about them (like a search engine). 
The directories structures are based on that of the LDAP ones, to solve scalability issues.

Both communicate through defined protocols (GRIP) for discovery (search) and enquiry (lookup). 
The schema followed is that of the LDAP one, using a hierarchical namespace, in order to aim at scalability issues.
The authors also define a notification mechanism (GRRP) for maintaining a soft-state of the resources.
The state may be discarded if it is not refreshed for some time, so the providers must implement a heartbeat.

%\paragraph{Notes}
The way an information provider knows whom it must register with is not clearly stated. The authors say manual configuration can be used,
but it would not be desirable in an autonomic management context. 
An alternative is to use another discovery service previously existent.
%In any cases, a hierarchical structure should be desirable.\\
%The autonomic management architecture described in Jade includes a {\em node discovery service}, used in the process of deploying a JadeNode (i.e. 
%recruiting a node to host an application module and apply the autonomic
%management requires to launch also a JadeNode on this host, also named node).
%The Jade {\em application adaption agent} is considered through the Jade reactors, which read the sensors, and apply changes through actuators.

\subsection{Autopilot: Adaptive Control of Distributed Applications}
\subsubsection{Paper reference}
R.L.Ribler, J.S.Vetter, H.Simitci, D.A.Reed,
Autopilot: Adaptive Control of Distributed Applications.
7th IEEE Int. Symposimum on High Performance Distributed Computing (HPDC-7), 1998.
\subsubsection{Brief summary of the project described in the paper}
This article describes a monitoring framework that is
further used  for dynamically adapting the behavior of a (distributed) application (see GrADS project \url{http://www.hipersoft.rice.edu/grads}), in
particular with the aim to maximise the performances of the application.

The framework includes:

   \begin{itemize}
      \item Distributed performance sensors, to monitor application and system performance, and generate qualitative and quantitative descriptions.
      \item Actuators, to modify the application behavior through the manipulation of parameter values.
      \item Distributed name servers, that works as registries of sensors and actuators: a client can then subscribe to a given sensor or actuator according
to the kind of monitoring information it is interested in (i.e. to read or
set a variable or parameter of the application)
      \item Decision mechanisms (like  Jade reactors are): they implement the algorithm that reads the information from the sensors, and decides actions to be implemented via the actuators. In the general case, this part is hard to accomplish.
%, but for testing purposes a very simplistic mechanism based on a table may be implemented
   \end{itemize}

Both sensors and actuators are represented through {\em property lists}, which are pairs property (or variable name)/value.
The events are represented in a particular data format.

\subsection{A Scalable Wide-Area Grid Resources Management Framework}
\subsubsection{Paper reference}
M. El-Darieby, D. Krishnamurthy,
 A Scalable Wide-Area Grid Resource Management Framework.
International conference on Networking and Services (ICNS'06).

\subsubsection{Brief summary of the project described in the paper}
This article describes a framework designed to attain scalable resource management.
The schema proposed is hierarchical, defining various levels of Resource Managers.

At the lowest levels, each resource has an Individual Resource Manager (IRM), each cluster has a Cluster Resource Manager (CRM).
Groups of cluster can be grouped in virtual clusters, with another CRM representing them, and constructing more levels of hierarchy.
In the top-level there are Grid Resource Managers (GRM) which manage resources for their lower levels, and also can communicate
with their peers (other GRMs) in order to submit jobs to appropiate locations.

%The information maintained by each manager is more abstract as  the level increases:
The more the level increases, the more the information maintained by each manager is abstract: 
   \begin{itemize}
      \item IRMs maintain information about resource state and availability.
      \item CRMs maintain aggregated information about the resources included in the managed cluster.
      \item CRMs at higher levels maintain information about the clusters managed.
      \item GRMs maintain summary information about the clusters.
   \end{itemize}

Lower level managers include more detailed information about the resources. 
This is done in order to avoid redundancy and improve scalability. 
Otherwise, GRMs would have a huge amount of information, and should have to maintain it up-to-date, too.

The assignments are propagated to the lower levels. 
Recoveries, and other actions that can be locally handled, need not to be 
propagated to higher levels.

\paragraph{Notes}
Although no implementation is presented, the ideas fit
well to our goal. Such hierarchical approach could
be taken in Jade to build a hierarchy of managers: next
part of the report presents 
a reorganisation of Jade following this direction.
%It's no clear how to organize the nodes to form the hierarchical structure. That should be done dynamically.

\subsection{A Grid Monitoring Architecture}
\subsubsection{Paper reference}
B. Tierny, R.~Aydt, D.~Gunter, W.~Smith, M.~Swany, V.~Taylor, R.~Wolski,
A Grid Monitoring Architecture.
Tech. Rep. GWD-PERF-16-2, Global
    Grid Forum, January 2002. \url{citeseer.ist.psu.edu/article/tierney02grid.html}.

\subsubsection{Brief summary of the project described in the paper}
This article describes the Grid Monitoring Architecture, developed by the Global Grid Forum Performance Working Group.

They list some considerations that have to be taken into account when designing a monitoring architecture:
   \begin{itemize}
      \item Performance data has a fixed, and often short lifetime utility. 
            Long-term storage is no needed, unless accounting is going to be done.
            Readings, on the other hand, have to be quick.
      \item Updates are frequent. Performance information is updated more frequently than it is read. 
            It is important to optimize updates, rather than queries.
      \item Performance information is often stochastic. Raw data may have to be processed in order to get relevant information.
   \end{itemize}
   
A grid monitoring architecture should meet the following requirements:
   \begin{itemize}
      \item Low latency in the transmission from sensors to consumers.
      \item High data rate transmission, as the performance information could be generated at a high rate.
      \item Minimal measurement overhead. 
      \item Secure. 
      \item Scalable. 
   \end{itemize}

The Grid Monitoring Architecture is built using 3 types of 'components':
   \begin{itemize}
      \item Directory Service: supports information publication, and discovery. May be distributed to improve scalability.
      \item Producer: makes performance data available (event source)
      \item Consumer: receives performance data (event sink)
   \end{itemize}

Producers and consumers register with the {\em Directory Service} (publish/suscribe interaction).
The {\em Directory Service} is used to locate producers and consumers, i.e. maintains registry of available producers and consumers, and answers queries about them.  After that, events flow directly between them. It does not store event data, that is, the information resulting from
monitoring. 
This allows the separation of Data Discovery and Data Transfer.

Consumers make queries to producers, requesting events (query/response interaction), and receives them. 
They can also make queries to the Directory Service, in order to locate appropriate producers.

Producers register with the Directory Service, accept queries from consumers and make the responses. They can also  notify events to subscribed consumers.

A component may implement one or both of the producer or consumer interfaces. This way, a component can act as producer, consumer, or both.

\paragraph{Notes}
In order to provide scalability, a set of components that implement both producer and consumer interfaces may be implemented.
In the context of Jade, those components could be the ones that serve
the role of sensor or actuator on the wrapped legacy application module.
For example such a Jade Fractal component could receive input from multiple producers through its consumer interface, aggregates it or derive some other metrics from them, and then send that result to another consumer through its producer interface. 
This way, the hierarchy could be built.

\subsection{A taxonomy of grid monitoring systems}
\subsubsection{Paper reference}
S.Zanikolas, R.Sakellariou,
A taxonomy of grid monitoring systems.
Future Generation Computer Systems 21 (2005), Elsevier
\subsubsection{Brief summary of the project described in the paper}
This paper presents a taxonomy for classifying grid monitoring systems. The criteria are taken from the 'components' implemented
in the system. 

Requirements they list for a Grid Monitoring System are:
   \begin{itemize}
      \item Scalability: Good performance on monitoring, and low intrusiveness on the monitored resources.
      \item Extensibility: Extensible data format, extensible producer-consumer protocol.
      \item Data delivery models: For example, measurement policies could be periodic or on-demand.
      \item Portability.
      \item Security.
   \end{itemize}

The 'components', by which the taxonomy is built, are based on those proposed by the GMA (Grid Monitoring Architecture), and include:
   \begin{itemize}
      \item Sensors. Processes that monitor an entity and generate events. Maybe merely passive, or make estimations (more intrusive).
      \item Producers. Processes that read data from sensors, and implement an interface to comunicate with directories (to register itself), 
and with consumers. Producers may also filter or summarize data.
      \item Republishers. Components that can implement both producer and consumer interface.
      \item Hierarchy of republishers. A structure containing one or more republishers hierarchically organized.
      \item Consumers. Processes that read data from producers through a defined interface.
   \end{itemize}

Additionally, there is another component called the Registry, which acts as a discovery service. 
Producers and consumers subscribe to it, in order to discover each other. 
It is also possible to request specific types of data, and, in this way, associate one consumer to the 'best' producer(s).

The taxonomy proposed is built around the numbers of components included in a monitoring system, and the systems are classified accordingly:
   \begin{itemize}
      \item Level 0 systems implement only Sensors and Consumers, which communicate in an application-specific way.
      \item Level 1 systems implement Sensors, Producers and Consumers, communicating through defined interfaces, allowing multiple consumers connecting to them.
      \item Level 2 systems implement Republishers instead of just Producers. Republishers may be centralized or distributed and support different functionalities.
      \item Level 3 systems implement Hierarchies of Republishers, which are reconfigurable, allowing for (potential) scalability.
   \end{itemize}
   
The cycle of monitoring encompasses the following phases, which can be accomplished by one or more different components of the system:
   \begin{itemize}
      \item Generation of events. At sensors or producers.
      \item Processing. At sensors, producers or republishers.
      \item Distribution. At producers, republishers, or consumers.
      \item Presentation and consumption. At consumers.
   \end{itemize}

\paragraph{Notes}
%The word 'components' is quite confusing since we also deal with a 'component' model. 
%Whenever possible I'll try to use another word, but in this case a 'component' is NOT a component-model component.
The architecture in which the taxonomy is based does not address the reconfiguration of the resources, nor some component that would be in charge
of 'reacting' to the monitoring. This is reasonable, as long as the focus is in monitoring, and not in autonomic management. If any, the autonomic management
algorithm to decide which actions have to be taken on the system, should be connected to the consumers.

\subsection{Synthesis and General remarks}

\begin{itemize}
\item The sensor process should be a low invasive process.

\item All collected information  will surely be outdated, as it changes dynamically. So, it doesn't make much sense to store large amounts of data, incurring in high overheads to maintain this data up-to-date.
  
\item As a consequence, it is necessary to transmit less amount of data through the monitoring entities, or well, use a particular format to make the processing easier.

\item The hierarchical approaches proposed in the various studied papers seem to provide the most promising basis for scalability. Different levels of the hierarchy may manage different kind of information. Higher levels may only receive summarized data, and lower levels (near to resources), manage the details.

\item In the context of the Jade system which is built upon the hierarchical Fractal software component model, it could be possible to build a hierarchy for sensors, and also for actuators.

\item The way to build the hierarchy of components in order to attain scalability should be done carefully and automatically at deployment time of the monitored application.
  The analysed works do not state how to do it, and some of them talk about a manually configured hierarchy. As the availability of the resources is dynamic, a manual configuration does not sound appropriate, even less in an autonomic managed context.
 
\end{itemize}

\newpage  
\section{Jade architecture}
\subsection{Presentation}
The Jade architecture is based on the Fractal component model. It is implemented in Java using Julia and Fractal RMI.
The main elements of the Jade architecture are the JadeBoot and  JadeNodes. 
A node (a machine) must host a JadeNode in order to be
managed by Jade. 
One node must host a JadeBoot, which is a component containing a set of 
components offering services that allow to
implement the autonomic behavior.

%Figure \ref{fig:jadeboot} describes the architecture of the JadeBoot component.

\begin{figure}[h]
\begin{center}
   \includegraphics[scale=0.50]{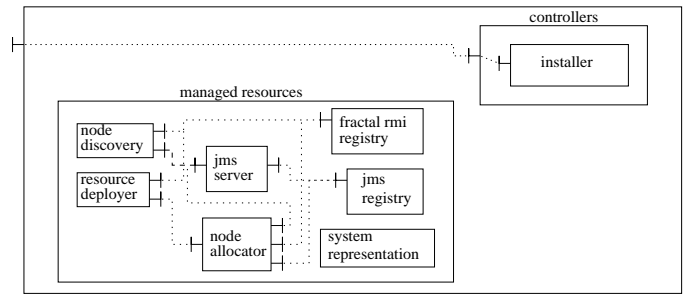}
\end{center}
\caption{Architecture of a JadeBoot}
\label{fig:jadeboot}
\end{figure}

As shown in Figure \ref{fig:jadeboot}, the JadeBoot component includes several {\em Common Services} in order to provide the autonomic behavior:

\begin{itemize}
   \item Node Discovery Service. Receives HeartBeats from a set of JadeNodes and keeps a dynamic list of available nodes.
         A node that takes too much (configurable) time to answer is marked as failed.
   \item Resource Deployer. Communicates with a remote installer and deploys and starts components containing a wrapped legacy application
         in a JadeNode provided by the Allocator component.
   \item Allocator. Manages a list of available JadeNodes, and keeps a mapping of deployed applications and nodes. 
         This component can implement an allocation policy in order to provide the 'best' node according to some criteria.
   \item JMS Server, and JMS Registry. Implements a messaging service to be read by other nodes and notifies events like node availability,
         or node failure.
   \item Fractal RMI registry. A naming service to locate components located on remote nodes.
   \item System Representation. A dynamically built component that reflects the current architecture of the JadeBoot and the JadeNodes.
         It is used to have a current state of the deployed applications and nodes, which can be useful to handle the failure of a node
	 containing a set of deployed applications.
\end{itemize}

%The architecture of the JadeNode component is described in Figure \ref{fig:jadenode}.

\begin{figure}
\begin{center}
   \includegraphics[scale=0.50]{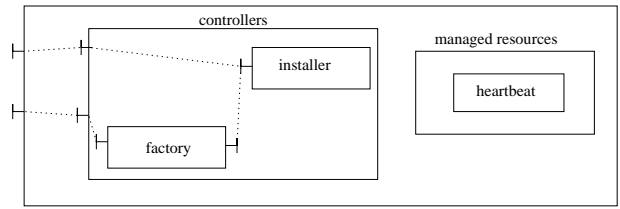}
\end{center}
\caption{Architecture of a JadeNode}
\label{fig:jadenode}
\end{figure}

The JadeNode component Figure (\ref {fig:jadenode}) includes a smaller set of components, and is able to install components containing wrapped legacy applications,
and make themselves available to a JadeBoot:

\begin{itemize}
   \item Factory. A GenericInstallingFactory to create new Fractal components containing wrapped legacy applications.
                  The application itself is deployed and installed as an OSGi bundle from a Bundle Repository, through a OSGiInstaller component.
   \item HeartBeat. Sends periodically messages to the JadeBoot in order to make itself available to the Jade architecture.
   
\end{itemize}

\subsection{Scalability issues}

%Part of my internship devoted some time to make work the Jade architecture inside a desktop cluster.

As the Jade architecture was developed to work inside a single cluster of machines providing some services, 
it lacks features  able to work over a more large-scale distributed infrastructure (e.g. a multi-cluster based grid or even a desktop grid).

\begin{itemize}
   \item The architecture comprises one JadeBoot controlling several JadeNodes. An approach that works well inside small clusters,
         but that is not scalable to a grid context.
   \item No automatic mechanism is provided to deploy the JadeBoot and JadeNodes on the cluster. 
         In fact, each component must be manually started on each node.
\end{itemize}

\newpage
\section{Scalable monitoring within Jade}

This section presents both an initial effort to deploy the Jade system using the ProActive deployment mechanism, and a proposition for a new architecture for Jade.

\subsection{ProActive-based deployment}

In order to deploy Jade nodes on several machines, we use an utility software named \textit{Command Launcher} based on ProActive Deployment Descriptors. The command Launcher  aims at executing a command (e.g. start  a software) on a given set of targets which are computing resources ``acquired'' through
a ProActive Deployment Descriptor. 
More precisely, the command launcher system deploys ProActive runtimes onto which Command Launcher objects (i.e. Active Objects) are launched and have the ability to execute the given command.

A ProActive Deployment Descriptor has been defined  (Figure \ref{ddex}), and can then be used (transparently to the user)
by the command launcher to automatically deploy a Jade architecture over a set of physical nodes. It is just a matter of modifying this descriptor 
in order to get access to an other grid
infrastructure or set of machines, according to the advocated job submission protocol. In the given example on Figure \ref{ddex}, we use ssh to get access
to hosts in the grid -- Grid 5000; so simply replacing, in the process definition, \texttt{ssh} by \texttt{oar} is sufficient to get access to the same grid but using a different job submission protocol.

%TODO : add a figure that shows the deployment descriptor used by Cristian 
%on the cluster or on the OASIS desktop 
\begin{figure}[h]
\begin{tabular}{@{}|c|@{}}
\hline
\\[-2ex]
\begin{minipage}{.95\linewidth}
\begin{scriptsize}
\begin{verbatim}
<?xml version="1.0" encoding="UTF-8"?>
<ProActiveDescriptor xmlns:xsi="http://www.w3.org/2001/XMLSchema-instance"
                     xsi:noNamespaceSchemaLocation="DescriptorSchema.xsd">
   
   <variables>
      <descriptorVariable name="NODES"  />
   </variables>
   
   <componentDefinition>
      <virtualNodesDefinition>
         <virtualNode name="grid"    timeout="${timeout}" property="multiple"/>
      </virtualNodesDefinition>
   </componentDefinition>
   <deployment>
      <mapping>
         <map virtualNode="grid">
            <jvmSet>
               <vmName value="g5k"/>
            </jvmSet>
         </map>
      </mapping>
      <jvms>
         <jvm name="g5k">
            <creation>
               <processReference refid="ssh_g5k_0"/>
            </creation>
         </jvm>
      </jvms>
      
   </deployment>
   
   <infrastructure>
      <processes>         
         <processDefinition id='g5k'>
            <processListbyHost class='org.objectweb.proactive.core.process.ssh.SSHProcessList' 
	                       hostlist='{$NODES}'>  
               <processReference refid='g5k'/>
            </processListbyHost>
         </processDefinition>
      </processes>
   </infrastructure>
</ProActiveDescriptor>
 
\end{verbatim}%$
\end{scriptsize}
\end{minipage}\\[-2ex]
\\
\hline
\end{tabular}
\caption{\label{ddex}The generic ProActive Deployment Descriptor used
  to deploy Jade nodes}
\end{figure}

%All nodes are deployed inside a ProActive runtime, using a Command Launcher, on several machines.  => TODO : clarify this, I do not understand
The syntax to use  the command launcher is given in  Figure \ref{cl}. We
also provide  as an example, the specific command line that the 
Jade user has simply to provide in order to trigger the remote installation of 
JadeNodes on some nodes of the grid. 
In this specific example, we explicitly name
the 3 hosts on which a command (here a script named jadeNode) has to
be executed. But command launcher can also be given a 
list of NODES in a less explicit way (e.g. just requesting 3 machines
of a given cluster or grid). 
Through such kind of command, on each machine an Oscar framework is launched, which installs and starts the JadeBoot or the JadeNode.
Once the JadeBoot and the JadeNode are deployed, a legacy application,
for example a J2EE application, can be deployed and managed from within a Jade architecture.

%an example are shown on : the user
%has just to specify a list of targets and the command to be executed.

\begin{figure}[h]
\begin{tabular}{@{}|c|@{}}
\hline
\\[-2ex]
\begin{minipage}{.95\linewidth}
\begin{scriptsize}
\begin{verbatim}


java CommandLauncher -DNODES={targets} command

java CommandLauncher -DNODES="sidonie.inria.fr meije.inria.fr naruto.inria.fr" jadeNode
\end{verbatim}
 \end{scriptsize}
\end{minipage}\\[-2ex]
\\
\hline
\end{tabular}

\caption{\label{cl}The use of Command Launcher}

\end{figure}

% (and so,
%decoupling the actions of getting access to grid machines,
%starting OSGi gateways, and then, installing Jade bundles). Once
%started, we could also think collecting system-level, OSGi gateway-level or
%Jade-level monitoring information taking advantage
%of such a typed group \cite{ICAS}.

%This could also allow to deploy the nodes as ProActive components.

%That architecture is specified in the ProActive deployment descriptor.
%Starting the JadeBoot and JadeNode remotely is easy, provided that both the JadeBoot and the JadeNode are available as OSGi Bundles. 

%The set of Command Launcher objects is handled as a ProActive typed group. 
%Consequently, such a group can  be useful serve to 
%start or stop Jade nodes  in a one-shot method invocation,
%and so in parallel. 

\subsection{Towards a hierarchical organisation of Jade}

\subsubsection{JadeMirror}
In the Jade architecture, monitors can be added as components. A monitoring cycle of Jade consists of Sensors, Reactors and Actuators.
Sensors can be located on the JadeNode (or the JadeBoot depending on what is going to be sensed), and transmit sensed data
to the Reactors. The Reactors implement the decision mechanism, based on the sensed information, and throw actions that are
going to be implemented through the Actuators. 

The Jade framework considers that several monitoring domains can coexist. For example, there can be a {\em repair} monitoring cycle
that will react when some node fails and will replace the failed node for another providing the same services; 
or an optimization cycle, that will modify some parameters
of the applications in order to optimize the response time, or the resources utilization.

For achieving these goals, the centralized approach where all the JadeNodes send the sensed data (through the sensors) to the 
JadeBoot, is not suitable. For that, we  are proposing to introduce an intermediate component, called {\em JadeMirror}.

The {\em JadeMirror} replicates the actions of the JadeBoot, and acts as a manager for all the nodes connected to it.

\subsubsection{JadeMirror design and associated implementation}

The architecture of both the JadeBoot and the JadeNode have been modified to include a Monitor component among their services. 
In the case of the JadeNode, it includes a LocalMonitor component, which is composed of a LocalSensor and a LocalActuator. 
As an example, the LocalSensor periodically reads CPU utilization, and sends a message to the JadeBoot. 
The JadeBoot also includes a Monitor component, which is composed of a Sensor, a Reactor and an Actuator, which behaves in
the previous described way. 
The Sensor receives the message from the LocalSensor in the JadeNode, and feeds the Reactor, which can take a decision
with that data and, possibly, call the Actuator to execute some action. 
That action could be executed directly by the Actuator (for example, if some binding between components has to be changed,
or if some node has to be stopped), or better by delegating the execution
of this action to the LocalActuator on the JadeNode 
(for example, if some parameter has to be tuned on the node in order to optimize some response).

\begin{figure}
\begin{center}
   \includegraphics[scale=0.50]{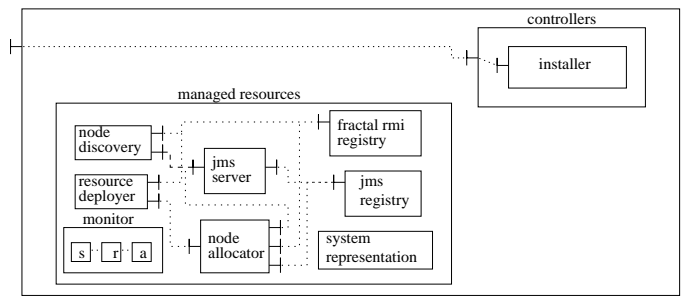}
\end{center}
\caption{Architecture of a JadeBoot featuring monitoring capabilities}
\label{fig:jadeboot2}
\end{figure}

\begin{figure}
\begin{center}
   \includegraphics[scale=0.50]{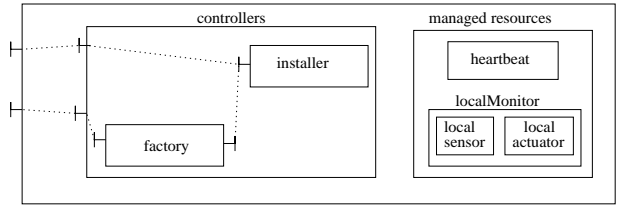}
\end{center}
\caption{Architecture of a JadeNode featuring monitoring capabilities}
\label{fig:jadenode2}
\end{figure}

This step  which has consisted in the addition of a monitoring part to the Jade architecture is depicted in Figures \ref{fig:jadeboot2} and 
\ref{fig:jadenode2}.

For achieving scalable monitoring, the addition of  JadeMirror  nodes is proposed. 
A JadeMirror includes almost the same components of a JadeBoot.
In particular, the Node Discovery Service, the Fractal RMI registry, JMS Server, JMS Registry and System Representation, 
can be included, offering  the same functionality as the JadeBoot.
The JadeMirror, as it must act like a JadeNode too, must include also the HeartBeat component. 
For the monitoring, the JadeMirror includes the LocalMonitor, to behave as any other JadeNode: it will send notifications
to a remote JadeBoot; but it  must also include a Monitor that receives data from its set of JadeNodes 
(here, the JadeMirror is acting like a JadeBoot).

Some components must be modified in order to behave correctly under the hierarchical scheme. 
For example, the Resource Deployer must try to allocate a node from a set of JadeNodes, but if there is no one available,
 it should forward the request to a higher level JadeBoot (or another JadeMirror higher in the hierarchy), letting it
to take the decision and possibly consulting other JadeMirrors managed by
some other JadeMirrors.

In the current state, the JadeMirror is implemented including the components of both a JadeBoot and a JadeNode. 
 The practical experiments we conducted actually comprised the deployment of 1 JadeBoot, 2 JadeMirrors, and 2 JadeNodes connected to each JadeMirror, using a total of 7 nodes.
In this scheme, the JadeNodes send information about CPU utilization to their "master" JadeMirror, and this one sends data to the central JadeBoot.

Our current experimental hierarchical Jade system does not yet include the 
necessary modifications to the Resource Deployer components, and Allocator:
that is, allowing them to send answers to their JadeBoot stating, for example, that there are no more available nodes
or that some action cannot be fulfilled and that a higher level node must take care of it.

\newpage
\section{Summary and next steps}

In this work, we have started to introduce some scalable monitoring capabilities in the Jade autonomic management system.
The implementation is not complete, but the current state it
has reached is sufficient to serve as a proof-of-concept about
the building of hierarchy of  Jade components (Boot, Mirror, Nodes). On 
this basis, we are expecting that scalable monitoring can be acheived.

The {\em gridification} of Jade is addressed by considering deploying the Jade architecture using ProActive deployment descriptors. 
Moreover, we are convinced that an implementation of 
Jade using 
the ProActive/GCM implementation of the Fractal component model 
\cite{gcm} could
improve the grid awareness of Jade: indeed, it would allow for 
interactions between the Fractal Jade components taking into account grid constraints
as high-latency, dynamicity, non-secure communication channels.

%\section{ref bibtex}

%czajkowski01grid \cite{czajkowski01grid}\\
%ribler98autopilot \cite{ribler98autopilot}\\
%eldiareby \cite{eldiareby}\\
%tierney02grid \cite{tierney02grid}\\
%serafeim \cite {serafeim}\\

%% \section{First section}
%% Here is some text for the first section, and a label\label{sec1}.
%% Uses version \RRfileversion\ of the package.\newpage
%% \section{Second section}
%% Text for the second section. This is closely related to the text in
%% section \ref{sec1} on page \pageref{sec1}. \newpage
\newpage
 \tableofcontents

\newpage 
\bibliographystyle{abbrv}
\bibliography{biblio}

\begin{thebibliography}{1}

\bibitem{BBCCHMQ05}
L.~Baduel, F.~Baude, D.~Caromel, A.~Contes, F.~Huet, M.~Morel, and R.~Quilici.
\newblock {\em {Grid Computing: Software Environments and Tools}}, chapter
  Programming, Composing, Deploying, for the Grid.
\newblock Springer Verlag, 2005.

\bibitem{BouchICAC05}
S.~Bouchenak, F.~Boyer, D.~Hagimont, S.~Krakowiak, N.~de~Palma, V.~Qu\'ema, and
  J.-B. StefaniD.
\newblock Architecture-based autonomous repair management: Application to
  {J2EE} clusters.
\newblock In {\em IEEE International Conference on Autonomic Computing
  (ICAC'05), short paper}, 2005.

\bibitem{Cluster}
S.~Bouchenak, N.~D. Palma, D.~Hagimont, and C.~Taton.
\newblock {Autonomic Management of Clustered Applications}.
\newblock In {\em IEEE International Conference on Cluster Computing (Cluster
  2006)}, Barcelona, Spain, Sept. 2006.

\bibitem{CDD:CMST06}
D.~Caromel, C.~Delb\'e, A.~di~Costanzo, and M.~Leyton.
\newblock {ProActive}: an integrated platform for programming and running
  applications on grids and {P2P} systems.
\newblock {\em Computational Methods in Science and Technology}, 12(1):69--77,
  2006.

\bibitem{gcm}
{CoreGRID Programming Model Virtual Institute}.
\newblock Basic features of the grid component model (assessed), 2006.
\newblock Deliverable of the CoreGRID Network of Excellence, Programming Model
  Virtual Institute, D.PM.04.

\end{thebibliography}

\end{document}